\documentclass{article}
\usepackage[utf8]{inputenc}
\usepackage{spconf,IEEEtrantools}
\usepackage{amsthm}
\newtheorem{theorem}{Theorem}

\newtheorem{lemma}[theorem]{Lemma}

\newtheorem{definition}{Definition}

\usepackage[english]{babel}
\usepackage{newtxtext, newtxmath}
\usepackage{cite}
\usepackage{graphicx}
\graphicspath{{Figures/}}
\usepackage[caption=false,font=footnotesize]{subfig}
\usepackage{cases}
\usepackage{mathtools}
\usepackage{xcolor}
\usepackage{url}
\usepackage{siunitx}
\usepackage{derivative}
\usepackage{booktabs}

\DeclareMathOperator{\bigO}{\mathcal{O}}

\DeclarePairedDelimiterX{\inner}[2]{\langle}{\rangle}{#1, #2}

\usepackage{ifthen}
\newcommand{\Oinf}[1]{
\ifthenelse{
	\equal{#1}{0}}
	{\mathcal{O}(1)}
	{\mathcal{O}(\varepsilon^{#1})}
}

\newcommand{\foo}[1]{%
  \ifthenelse{\equal{\detokenize{#1}}{\detokenize{german}}}
    {TRUE}
    {FALSE}%
}

\newcommand{\tpwm}{T_{\rm pwm}}

\def\BR{{\mathbb R}}

\title{Error estimates in Second-order Continuous-Time Sigma-Delta modulators}
\name{Dilshad Surroop$^{1,2}$ \qquad Pascal Combes$^{2}$ \qquad Philippe Martin$^{1}$}

\address{$^{1}$Centre Automatique et Systèmes, MINES ParisTech, PSL University, Paris, France \\
			    $^{2}$ Industrial Automation Business, Schneider Electric, Pacy-sur-Eure, France}
\begin{document}
%
\maketitle

\begin{abstract}
	Continuous-time Sigma-Delta (CT-$\Sigma\Delta$) modulators are oversampling Analog-to-Digital converters that may provide higher sampling rates and lower power consumption than their discrete counterpart. Whereas approximation errors are established for high-order discrete time $\Sigma\Delta$ modulators, theoretical analysis of the error between the filtered output and the input remain scarce. This paper presents a general framework to study this error: under regularity assumptions on the input and the filtering kernel, we prove for a second-order CT-$\Sigma\Delta$ that the error estimate may be in $o(1/N^2)$, where $N$ is the oversampling ratio. The whole theory is validated by numerical experiments. 
\end{abstract}
\begin{keywords}
	Sigma-Delta modulator, Continuous, Analog-to-Digital conversion (ADC), Approximation
\end{keywords}

\section{Introduction}

Introduced by Inose and Yasude \cite{InoseY1963IEEE}, Sigma-Delta ($\Sigma\Delta$) modulators are nowadays widely used Analog-to-Digital converters. Such 1-bit ADCs operate at many times the Nyquist rate, and can achieve the same resolution as Nyquist ADCs with suitable signal processing~\cite{SchreT2005book}. For kth-order discrete-time $\Sigma\Delta$ modulators, works by Daubechies, Güntürk and al. \cite{DaubeD2003AoM, Guntu2003CPAM, GuntuT2004IEEE} provide estimates for the error between the filtered output and the input. Typically, they obtain a mean-squared error estimate in $\bigO(1/N^k)$ for a time-varying input, where $N$ is the oversampling ratio, using for example a $\operatorname{sinc}^{k+1}$ filter \cite{Thao2001ISCAS}. But for continuous-time $\Sigma\Delta$ (CT-$\Sigma\Delta$) modulators, such general results remain partial. These CT-$\Sigma\Delta$ modulators deliver more power-efficient operations than their discrete-time equivalent, as well as higher sampling rates \cite{OrtmaG2006book,SchreT2005book, StraaP2008JSSC}.

Privileged in high performance motor control \cite{Soren2015PCIM}, the $\Sigma\Delta$ ADC is used to retrieve the phase currents which carry information on the rotor position if correctly filtered \cite{SurroCMR2020arxivb}. Indeed, the Pulse-Width Modulation (PWM) of the input voltage creates ripples in the current measurements \cite{SurroCMR2020ACC} that we can extract through a demodulation procedure using linear combination of iterated moving averages \cite{SurroCMR2019IECON}. Therefore, knowing the error estimate of the $\Sigma\Delta$ modulator is of utmost importance for this type of application. 

We present a general technique to study higher-order CT-$\Sigma\Delta$ modulators. Under regularity assumptions on the input and the filtering kernel, we prove for a second-order CT-$\Sigma\Delta$ that the error estimate may be in $o(1/N^2)$. 

This paper is organized as follows: we first detail the required definitions and technical lemmas; then we prove the error estimate on a specific second-order $\Sigma\Delta$ modulator. The theory is finally validated on numerical examples.

\section{Error estimate for a CT-$\Sigma\Delta$ modulator}
\label{sec:theory}
\subsection{Notations, definitions, preliminary results}
We consider the second-order CT-$\Sigma\Delta$ modulator depicted in figure~\ref{fig:sigma-delta}; $u$ denotes the input of the modulator which varies in a timescale $1/\tpwm$, $\nu \in \lbrace{0,1\rbrace}$ its output, $T_s$ its the sampling time, $x_{1,2}$ the states of the modulator, $N := \tpwm/T_s$ the oversampling ratio. We assume the stability of the modulator, which means both $x_1$ and $x_2$ are bounded. 

The notation $\mathcal{O}$ denotes the ``big O'' of analysis, i.e. $f(t,\varepsilon) = \Oinf{}$ if there exists $K >0$ independent of $t$ and $\varepsilon$ such that $\|f(t,\varepsilon)\| \leq K\varepsilon$. Likewise, the notation $o$ is the ``small o'' of analysis, i.e. $f(t,\varepsilon) = o(\varepsilon)$ if $\|f(t,\varepsilon)\| \leq \varepsilon g(\varepsilon)$ where $\lim_{\varepsilon\to 0} g(\varepsilon) = 0$.

The proof in subsection~\ref{subsec:sigma-delta} relies on the application of a generalization of the classical Riemann-Lebesgue lemma:

\begin{lemma}[Generalized Riemann-Lebesgue lemma \cite{Kahan1980}]
	\label{th:riemann_lebesgue}
	Let $\beta \in L^\infty [0,+\infty) $ such that $\beta$ has a finite mean value $\overline\beta$, with
	\begin{IEEEeqnarray*}{rCl}
		\overline\beta := \lim_{T\to+\infty} \frac{1}{T} \int_0^T \beta(t)\, dt.
	\end{IEEEeqnarray*}
	Then for every $f \in L^1 [0,+\infty)$, 
	\begin{IEEEeqnarray*}{rCl}
		\lim_{N\to+\infty} \int_0^{+\infty} \beta(Nt) f(t) \, dt = \overline\beta \int_0^{+\infty} f(t) \, dt. 
	\end{IEEEeqnarray*}
\end{lemma}

In the sequel, we will assume the input~$u$ to the modulator is $AC^1$, or possibly only piecewise $AC^1$, as defined below. This (rather modest) requirement is motivated by the fact that we need to use integration by parts on its derivative, see lemma~\ref{lemma:IPP}.
\begin{definition}[$AC^p$ functions]
	A function $f : I \subset \BR \to \BR$ is $AC^p$ on an interval $I$ if it is p-times differentiable and its pth-order derivative $f^{(p)}$ is absolutely continuous. It is piecewise $AC^p$ if $f$ is p-times differentiable and $f^{(p)}$ is piecewise absolutely continuous. 
\end{definition}

\begin{lemma}[Integration by parts for piecewise $AC^0$ functions]\label{lemma:IPP}
	\label{lemma:integration_by_parts}
	Consider $f \in L^1[a,b]$ with $-\infty \leq a < b \leq +\infty$, $F$ a primitive of $f$, and $g$ a piecewise $AC^0$ function. Set $I := \cup_{0\leq i\leq m} [x_i, x_{i+1}]$, with $a = x_0 < x_1 < \ldots < x_m = b$, such that $g$ is $AC^0$ on each $[x_i, x_{i+1}]$; as $g$ is piecewise $AC^0$, it is differentiable almost everywhere, with $g'$ as derivative. Then
\begin{IEEEeqnarray*}{rCl}
	\int_a^b f(\sigma)g(\sigma) \, d\sigma &=& \sum_{i=0}^{m-1} \bigl[ F(x_{i+1}^-)g(x_{i+1}^-) - F(x_{i}^+)g(x_{i}^+) \bigr] \\
	&&\,- \int_a^b F(\sigma)g'(\sigma) \, d\sigma.
\end{IEEEeqnarray*}
\end{lemma}

\subsection{Second-order CT-$\pmb{\Sigma\Delta}$ ADC}
\label{subsec:sigma-delta}
\begin{figure}
	\centering
	\includegraphics[scale=0.9]{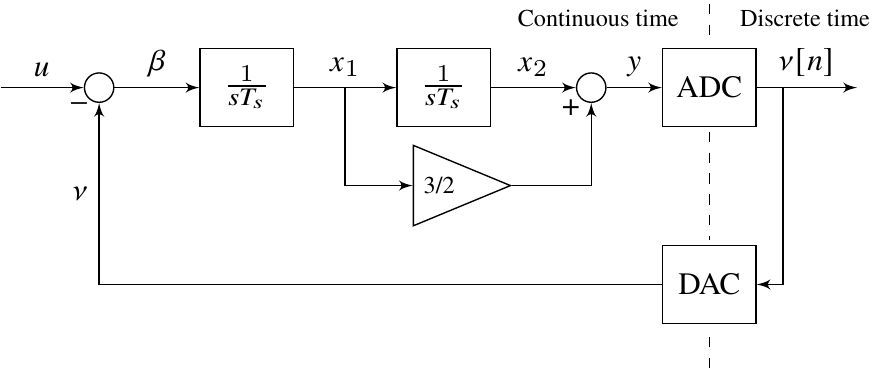}
	\caption{Example of second-order $\Sigma\Delta$ modulator \cite{SchreT2005book}}
	\label{fig:sigma-delta}
\end{figure}

We consider the modulator depicted in figure~\ref{fig:sigma-delta}. Its behavior is described by
\begin{IEEEeqnarray*}{rCl}
	T_s \dot x_1(t) &=& u\bigl(t/\tpwm\bigr) - \nu\bigl(t/T_s\bigr) \\
	T_s \dot x_2(t) &=& x_1(t) \\
	y(t)  &=& x_2(t) + \tfrac{3}{2}x_1(t).
\end{IEEEeqnarray*}
In the normalized time $\tau := t/\tpwm$, this becomes
\begin{subequations}
\begin{IEEEeqnarray}{rCl}
	\label{subeq:x1}
	\frac{1}{N} \dot x_1(\tau) &=& u(\tau) - \nu(N\tau)  \\
	\label{subeq:x2}
	\frac{1}{N} \dot x_2(\tau) &=& x_1(\tau) \\
	y(\tau)  &=& x_2(\tau) + \tfrac{3}{2}x_1(\tau)
\end{IEEEeqnarray}
\end{subequations}

We first prove that $\beta(N\tau) := u(\tau) - \nu(N\tau)$ admits a zero-mean primitive $\beta^{(-1)}$, which also has a zero-mean primitive $\beta^{(-2)}$. Integrating \eqref{subeq:x1} from $0$ to $t$ yields
\begin{IEEEeqnarray*}{rCl}
	\frac{1}{Nt}\bigl(x_1(t) - x_1(0)\bigr) &=& \frac{1}{t} \int_0^t u(\sigma) \, d\sigma -  \frac{1}{t} \int_0^t \nu(N\sigma) \, d\sigma.
\end{IEEEeqnarray*}
The modulator is assumed to be stable, so $x_1$ is bounded; the left-hand side of the previous equation vanishes when $t$ tends to infinity, and
\begin{IEEEeqnarray*}{rCl}
	\lim_{t\to+\infty} \frac{1}{t} \int_0^t \bigl[ u(\sigma) - \nu(N\sigma) \bigr] \, d\sigma = 0
\end{IEEEeqnarray*}
i.e., by definition, $\overline \beta = 0$. Integrating \eqref{subeq:x2} from $0$ to $t$ yields
\begin{IEEEeqnarray*}{rCl}
	\frac{1}{Nt}(x_2(t) - x_2(0)) = \frac{1}{t} \int_0^t x_1(\sigma) \, d\sigma
\end{IEEEeqnarray*}
Since $x_2$ is bounded as we consider the modulator is stable,
\begin{IEEEeqnarray*}{rCl}
	\overline x_1 = \lim_{t\to+\infty} \frac{1}{t} \int_0^t x_1(\sigma) \,d\sigma = 0.
\end{IEEEeqnarray*}
So $\frac{1}{N}x_1(\tau)$ has zero mean, and by~\eqref{subeq:x1}, it is the primitive of $\beta(N\tau)$. Thus $\beta^{(-1)}(N\tau) := \frac{1}{N} x_1(\tau)$ is the zero-mean primitive of $\beta(N\tau)$. Now integrating equation~\eqref{subeq:x2} from $0$ to $t$ gives
\begin{IEEEeqnarray*}{rCl}
	\frac{1}{N^2} \bigl( x_2(t) - x_2(0) \bigr) &=& \int_0^t \frac{1}{N} x_1 (\sigma) \, d\sigma  = \int_0^t \beta^{(-1)} (N\sigma) \, d\sigma 
\end{IEEEeqnarray*}
The left-hand side is bounded, so every primitive of $\beta^{(-1)}$ is bounded as well. Consequently, $\beta^{(-2)}$, the zero-mean primitive of $\beta^{(-1)}$ is well-defined.

\subsection{Filtering process}
\begin{figure}
	\centering
	\includegraphics{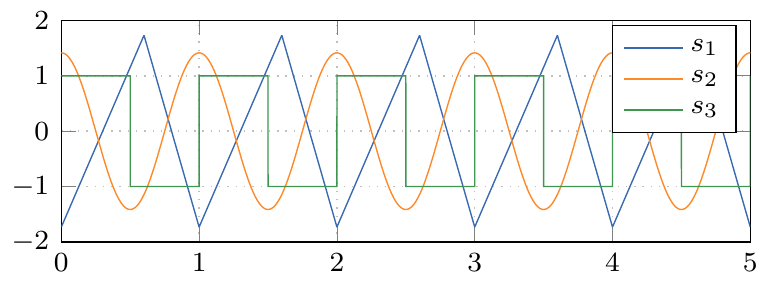}
	\caption{Signals $s_1$, $s_2$ and $s_3$}
	\label{fig:s}
\end{figure}

Theorem~\ref{th:filtering} provides an estimate for functions $\beta$ such that $\beta^{(-2)}$ and $\beta^{(-1)}$ with zero mean exist. 
\begin{theorem}
	\label{th:filtering}
	Consider $\beta \in L^\infty[0,+\infty)$ such that the zero-mean primitive $\beta^{(-1)}$ of $\beta$ exists, as well as the zero-mean primitive $\beta^{(-2)}$ of $\beta^{(-1)}$. Consider as well $K^k$ a twice differentiable kernel with support in $[0,k]$, and such that $K^k(0) = K^k(k) = (K^{k})'(0) = (K^{k})'(k) = 0$. 

	If $s$ is $AC^1$, then for $t \geq 0$,
\begin{IEEEeqnarray*}{rCl}
	I(t) := \int_\BR \beta(N\sigma) s(\sigma)  K^k_t (\sigma) \, d\sigma = o(1/N^2),
\end{IEEEeqnarray*}
with $K^k_t(\sigma) = K^k (t-\sigma)$. If $s$ is only piecewise $AC^1$, then for $t\geq 0$, $I(t) = \bigO(1/N^2)$.
\end{theorem}
In other words, the instantaneous difference between the filtered input and the filtered output is in $o(1/N^2)$ under some regularity assumptions on the kernel $K^k$ and the input $u$. 

\begin{figure}
	\centering
	\includegraphics{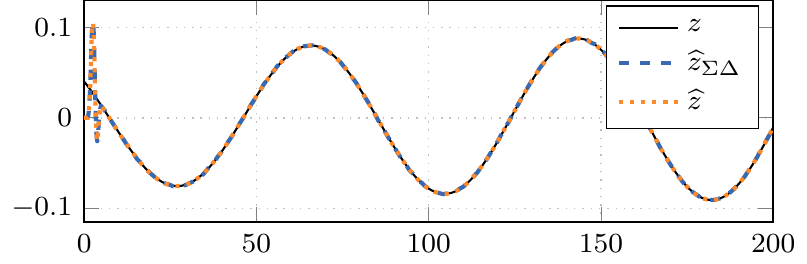}
	\hspace*{0.8em}\includegraphics{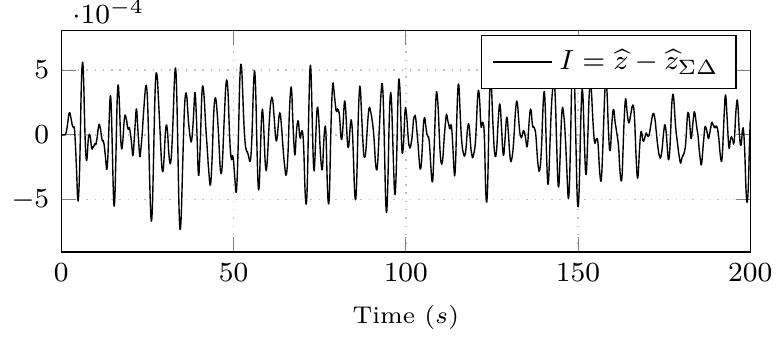}
	\caption{Signal $z$, estimates $\widehat z_{\Sigma\Delta}$ and $\widehat z$ (top), error $I(t) = \widehat z - \widehat z_{\Sigma\Delta}$ (bottom) for the input $u_1 = zs_1$. }
	\label{fig:z}
\end{figure}

\begin{proof}
	If $s$ is $AC^1$ (resp. piecewise $AC^1$), then $f_t : \sigma \mapsto s(\sigma)K^k_t(\sigma)$ is also $AC^1$ (resp. piecewise $AC^1$). In any case, $f_t$ is differentiable with support $[t-k,t]$ and a basic integration by parts gives
\begin{IEEEeqnarray*}{rCl}
	I(t) &=& \frac{1}{N} \bigl[ \beta^{(-1)}(Nt)f_t(t) - \beta^{(-1)}(N(t-k))f_t(t-k) \bigr]
	\\
	&&\, - \frac{1}{N}\int_{t-k}^t \beta^{(-1)}(N\sigma) f_t'(\sigma) \, d\sigma,
\end{IEEEeqnarray*}
where the first term is zero since $f_t(t) = f_t(t-k) = 0$.

We write $t- k = \sigma_0 < \ldots < \sigma_{m} =  t$ the locations of the loss of regularity of $s$. The integration by parts, given by lemma~\ref{lemma:integration_by_parts}, yields
\begin{IEEEeqnarray}{rCl}
	I(t) &=& -\frac{1}{N^2}\sum_{i=0}^{m-1} \bigl[\beta^{(-2)}(N\sigma_{i+1}^-) f_t'(\sigma_{i+1}^-) -\beta^{(-2)}(N\sigma_{i}^+) f_t'(\sigma_{i}^+)  \bigr] \notag\\
	     &&\, + \frac{1}{N^2}\int_{t-k}^t \beta^{(-2)}(N\sigma) f_t''(\sigma) \, d\sigma.
	\label{eq:I2}
\end{IEEEeqnarray}
The limit of the integral term in~\eqref{eq:I2}, by lemma~\ref{th:riemann_lebesgue}, is
\begin{IEEEeqnarray*}{rCl}
	\lim_{t\to+\infty}\int_{0}^{+\infty}\hspace{-0.7em} \beta^{(-2)}(N\sigma) f_t''(\sigma) \, d\sigma = \overline{\beta^{(-2)}} \int_{0}^{+\infty}\hspace{-0.7em} f_t''(\sigma) \, d\sigma  = 0,
\end{IEEEeqnarray*}
i.e. $\frac{1}{N^2}\int_{t-k}^t \beta^{(-2)}(N\sigma) f_t''(\sigma) \, d\sigma = o(1/N^2)$.  If $f$ is $AC^1$, the sum in~\eqref{eq:I2} is zero since $f_t'(t) = f_t'(t-k)$; therefore $I(t) = o(1/N^2)$. If $f$ is only piecewise $AC^1$, the sum in~\eqref{eq:I2} is not necessarily zero, and
$I(t) = \bigO(1/N^2)$, which concludes the proof. 
\end{proof}

\section{Numerical results}
\begin{figure}
	\centering
	\includegraphics{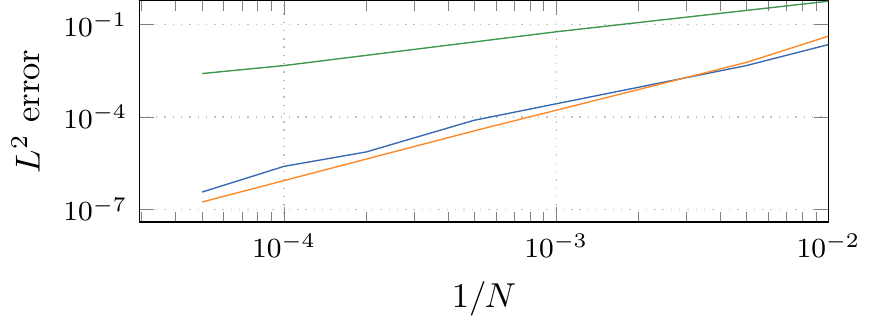}
\caption{Asymptotic behavior of $\|I\|_2$ as a function of $1/N$ for $u_1$ (blue, slope $\simeq 2$), $u_2$  (orange, slope $\simeq 2.3$) and $u_3$ (green, slope $\simeq 1$) }
\label{fig:asymp}
\end{figure}
The estimates obtained in section~\ref{sec:theory} are now validated on a numerical example. We consider the modulator of figure~\ref{fig:sigma-delta}, with $T_s = 5\times 10^{-3}$ \si{\second}. The tests are conducted with three different inputs $u_i(t) := z(t)s_i(t)$, with  $z(t) := 0.04\cos\bigl(\tfrac{t}{12}\bigr)-0.06\sin\bigl(\tfrac{t}{4\pi}\bigr)$, and
\begin{IEEEeqnarray*}{rCl}
	s_1(t) &:=& \frac{1}{\sqrt{0.03}}\left(\tau 1_{[0,0.6]}(\tau) + 1.5(1-\tau)1_{]0.6,1]}(\tau) - 0.3\right), \\
	s_2(t) &:=& \sqrt{2} \cos (2\pi \tau), \qquad s_3(t) := 1_{[0,0.5]}(\tau) - 1_{]0.5,1]}(\tau),
\end{IEEEeqnarray*}
where $\tau = \operatorname{mod}(t,\tpwm)/\tpwm$, $\tpwm = 1$ and $t \in [0,250]$.
Illustrated in figure~\ref{fig:s}, the $s_i$'s are respectively piecewise $AC^1$ ($s_1$), $AC^1$ ($s_2$) and discontinuous ($s_3$), and such that $\| s_i \|_2 = 1$. A kernel satisfying the hypotheses of theorem~\ref{th:filtering} is the convolution power of the characteristic function $1_{[0,1]}$,  $K^3 := 1_{[0,1]} * 1_{[0,1]} * 1_{[0,1]}$, as $\operatorname{supp}K^3 = [0,3]$ and $K^3(0) = K^3(3) = (K^{3})'(0) = (K^{3})'(3) = 0$ (see for example \cite{Aubin2011Wiley}); this kernel corresponds to a triple moving average.  

Define $\widehat z$ (resp $\widehat z_{\Sigma\Delta}$) the filtered input (resp. output) as
\begin{IEEEeqnarray*}{rCl}
	\widehat z(t) &:=& \int_0^{+\infty} u(\sigma) s(\sigma) K^3(t-\sigma) \, d\sigma \\
	\widehat z_{\Sigma\Delta}(t) &:=& \int_0^{+\infty} \nu(\sigma) s(\sigma) K^3(t-\sigma) \, d\sigma,
\end{IEEEeqnarray*}
so that $I(t) = \widehat z(t) - \widehat z_{\Sigma\Delta}(t) $. The estimates $\widehat z$ and $\widehat z_{\Sigma\Delta}$ are illustrated in figure~\ref{fig:z}, as well as their difference $I(t)=\widehat z - \widehat z_{\Sigma\Delta}$: $I(t)$ is as anticipated small, which shows the commutation of the filtering process with the $\Sigma\Delta$ modulator. 

To confirm the asymptotic behavior described by theorem~\ref{th:filtering}, the same simulation is carried out for each input $u_i$ and different values of $N$; for each experiment  the $L^2$-error $\|I\|_2 := (\int_1^{250} I(\sigma)^2 \, d\sigma))^{1/2}$ is computed. Figure~\ref{fig:asymp} shows these behaviors for the three inputs $u_i$ and validates the approximation orders. Indeed, when $s = s_1$ is piecewise $AC^1$, the approximation order is in $\bigO(1/N^2)$; it is slightly better when $s = s_1$ is $AC^1$, with $\|I\|_2 = \bigO(1/N^{2.3}) = o(1/N^2)$;  when  $s = s_3$ is discontinuous, we only have an estimate in $\bigO(1/N)$.

\section{Conclusion}
Depending on the regularity of the input, and assuming the modulator is stable, we proved the error between the filtered output and filtered input decreases at a rate which is $o(1/N^2)$ if the input is differentiable with a derivative that is absolutely continuous. Such an approximation error is crucial for some applications, for instance for current ripple extraction in sensorless control of electric motors.

\bibliographystyle{IEEEbib}
\bibliography{biblio.bib}

%
%

\end{document}